Hydrostatic pressure study of single-crystalline UNi$_{0.5}$Sb$_2$

B. K. Davis,[1,2] M. S. Torikachvili,[1] E. D. Mun,[3] J. C. Frederick,[3] G. J. Miller,[4] S. Thimmaiah,[4] S. L. Bud'ko,[3] and P. C. Canfield[3]

[1] Department of Physics, San Diego State University, San Diego, CA 92182-1233
[2] Quantum Design, 6325 Lusk Boulevard, San Diego, CA 92121
[3] Ames Laboratory and Department of Physics and Astronomy, Iowa State University, Ames, IA 50011
[4] Ames Laboratory and Department of Chemistry, Iowa State University, Ames, IA 50011

Abstract

We studied single-crystals of the antiferromagnetic compound UNi$_{0.5}$Sb$_2$ ($T_N \approx 161$ K) by means of measurements of magnetic susceptibility ($\chi$), specific heat ($C_p$), and electrical resistivity ($\rho$) at ambient pressure, and resistivity under hydrostatic pressures up to 20 kbar, in the temperature range from 1.9 to 300 K. The thermal coefficient of the electrical resistivity ($d\rho/dT$) changes drastically from positive below $T_N$ to negative above, reflecting the loss of spin-disorder scattering in the ordered phase. Two small features in the $\rho$ vs $T$ data centered near 40 and 85 K correlate well in temperature with features in the magnetic susceptibility and are consistent with other data in the literature. These features are quite hysteretic in temperature, i.e., the difference between the warming and cooling cycles are about 10 and 6 K, respectively. The effect of pressure is to raise $T_N$ at the approximate rate of 0.76 K/kbar, while progressively suppressing the amplitude of the small features in $\rho$ vs $T$ at lower temperatures and increasing the thermal hysteresis.

The intermetallic ternary compounds with general composition U$T$Sb$_2$ ($T$ = transition metal) form a large family of materials, most of which crystallize in a simple tetragonal HfCuSi$_2$-type structure (P4/nmm, No. 129).[1] Planar layers of Sb, $T$, and U-Sb are stacked along the c-axis, conferring these materials with strongly anisotropic properties. The U$T$Sb$_2$ compounds were found to order ferromagnetically (FM) for $T$ = Co, Cu, Ag, and Au, and antiferromagnetically (AFM) for $T$ = Ni, Ru, and Pd, in temperatures below 300 K.[1]

Attempts of growing single-crystals of UNiSb$_2$ from Sb flux resulted in crystals with a 0.5 occupancy at the Ni-site, yielding UNi$_{0.5}$Sb$_2$ as the actual composition.[2] The temperature ($T$) dependence of the electrical resistivity ($\rho$) and magnetic susceptibility ($\chi$) are reminiscent of other anisotropic AFM materials. The behavior of $\chi(T)$ suggests that the c-axis is the easy axis for magnetization, and that the Néel temperature ($T_N$) is ≈ 161 K.[2,3] The $\rho(T)$ data show a drop near $T_N$, consistent with a loss in spin-disorder scattering in the ordered phase. However, in addition to the features in $\rho(T)$ and $\chi(T)$ near $T_N$, two much smaller features centered near 40 K and 85 K could be detected.[2,3]

In order to probe the magnetic properties of UNi$_{0.5}$Sb$_2$, and to try to understand the origin of the 2 small features in $\rho(T)$ and $\chi(T)$ much below $T_N$, we carried out measurements of in-plane $\rho(T)$ in pressures up to 20 kbar, as well as measurements of magnetic susceptibility and specific heat. The single-crystals for this work were grown from a Sb-rich flux, using a technique described in Ref. 4. Powder XRD analysis showed that these crystals are single-phase. However, an intensity analysis of the XRD data from a single-crystal diffractometer revealed that the occupancy at the Ni site is 0.5 ± 0.01, confirming that the correct composition is UNi$_{0.5}$Sb$_2$. Similar widths of formation have been seen in many $RT$Sb$_2$ compounds ($R$ = rare earth, $T$ = Mn, Fe, Co, Ni, Cu, Zn, and Cd; see Ref. 4 and references cited therein for details). Measurements of magnetic susceptibility, magnetization, heat capacity, and electrical resistivity were carried out with the vibration sample magnetometer, calorimeter, and dc resistivity options of a Quantum Design PPMS-9, respectively. Measurements of $\rho(T,P)$ in hydrostatic pressures up to 20 kbar were carried out using a self-clamping piston-cylinder Be-Cu pressure cell with a hardened NiCrAl-alloy core, which was fit to the PPMS. Four Pt leads were attached to the sample using Epotek H20E Ag-loaded epoxy. The sample leads, and coils of manganin and Pb, which served as manometers, were attached to 12 Cu wires at the end of a Stycast-sealed feedthrough. This assembly was inserted into a Teflon cup filled with a 40:60 mixture of mineral oil:n-pentane, which served as the pressure transmitting medium. The pressure at room temperature was determined from the change in resistance of the calibrated manganin coil, while the pressure at low temperatures was determined from the change in the superconducting transition temperature of Pb. In light of the change in pressure in this type of cell upon cooling,[5] reflecting the different thermal expansion characteristics of the cell body and the pressure transmitting medium, we estimated the pressure values between 1.9 and 300 K by assuming a linear change of $P$ between 300 K and 90 K, and neglecting any $P$ changes for $T < 90$ K. The cooling and warming rates of the cell were kept close to 0.25 K/min, and the sample temperature was inferred from a cernox sensor placed on the body of the cell. The difference in sample temperature between the cooling and warming cycles was negligible.

Magnetization ($M$) curves in temperatures between 2 and 300 K showed a linear behavior of $M$ vs $H$ (data not shown). The behavior of $\chi^{-1}$ vs $T$ ($\mu_0 H = 1$ T) for $H$ parallel and perpendicular to the easy axis (c-axis) are shown in Figure 1. These data clearly show the onset of AFM order near 161 K. A fit of the $\chi^{-1}(T)$ data for $T > 250$ K to a Curie-Weiss expression $\chi = \chi_0 + C/(T-\theta)$, yielded the effective moments of $\mu_{eff} \approx 3.15$ ($H//c$) and 3.25 $\mu_B$ ($H \perp c$), somewhat reduced from the $U^{3+}$ and $U^{4+}$ values of 3.58 and 3.62 $\mu_B$, respectively. The two small features in the $\chi(T)$ data centered near 40 K and 85 K are hysteretic, and the higher values of $\chi$ are found in the cooling cycle, as indicated in the inset of Fig. 1.

The $C_p(T)$ data display a pronounced peak near 161 K, as shown in Fig. 2, consistent with the onset of AFM order. An extrapolation of the $C/T$ vs $T^2$ data at low temperatures to $T = 0$ yields a Sommerfeld coefficient $\gamma = 9.5$ mJ/mole·K$^2$ (Fig. 2 inset).

The curves of normalized electrical resistivity $\rho/\rho_{300K}$ vs $T$ for various pressures are shown in Fig. 3. The data for $P > 0$ are offset for clarity. The $\rho(T)$ data for $P = 0$ show two distinct features; 1) the temperature coefficient of the electrical resistivity, which is negative in the paramagnetic phase, becomes positive below $T_N$, reflecting a loss in spin-disorder scattering; 2) the two small features centered near 40 and 85 K are both hysteretic. These later 2 features remain unaffected by a 9 T magnetic field ($H//c$-axis; data not shown). The effect of pressure on the onset of AFM order is to raise the value of $T_N$ at the rate of $\approx 0.76$ K/kbar, as shown in the Fig. 3 inset. In addition, pressure first lowers the onset temperature on cooling of the two low temperatures features, while broadening the width of the hysteretic regions. The $\rho$ vs $T$ data show that the 2 features start to overlap for $P_{300K} = 15.4$ kbar ($P_{T<90K} \approx 14.8$ kbar). Eventually the two features are completely suppressed under pressure, as shown in the $\rho$ vs $T$ data for $P_{300K}$ = 17.8 kbar ($P_{T<90K}$ ≈ 20.6 kbar), and the data for the cooling and warming cycles cannot be distinguished.

The two small features in $\rho(T)$ and $\chi(T)$ centered near 40 K and 85 K correlate well in temperature, and the distinction between the warming and cooling cycles suggests that they are due to first order phase transitions. It is somewhat puzzling that these anomalies were not reflected in the $C_p(T)$ data, which were taken on the same sample used for the $\rho(T)$ and $\chi(T)$ measurements. Also, XRD data at 10 and 300 K don't reveal any crystallographic distortions at low temperatures.[2] The cooling branch of $\chi(T)$ corresponds to the higher values of $\chi$. On the other hand, the cooling branch of the $\rho(T)$ data corresponds to the lower values of $\rho$ near 85 K, and higher values near 40 K. It is tempting to assume that the two low temperature features are due to spin realignment within the AFM ordered phase. If this is the case, the effect of hydrostatic pressure is to stabilize the high-temperature AFM phase to lower temperatures, completely suppressing the spin realignment transitions in pressures near 20 kbar. In order to elucidate whether the two low temperature are in fact due to spin realignment, neutron scattering studies both in ambient and high pressure are in order.

The support from NSF Grant No. DMR-0306165 for the work at SDSU, and from the USDOE Contract DE-AC02-07CH11358 for work at Ames Laboratory and ISU are gratefully acknowledged.

References

[1] D. Kaczorowski, R. Kruk, J. P. Sanchez, B. Malaman, and F. Wastin, Phys. Rev. B **58,** 9227 (1998).
[2] Z. Bukowski, D. Kaczorowski, J. Stepien-Damm, D. Badurski, and R. Troc, Intermetallics **12,** 1381 (2004).
[3] S. Ikeda, T. D. Matsuda, A. Galatanu, E. Yamamoto, Y. Haga, and Y. Onuki, J. Mag. Mag. Mat. **272-276,** 62 (2004).
[4] K. D. Myers, S. L. Bud'ko, I. R. Fisher, Z. Islam, H. Kleinke, A. H. Lacerda, and P. C. Canfield, J. Magn. Magn. Mat. **205,** 27 (1999).
[5] J. D. Thompson, Rev. Sci. Instrum. **55,** 231 (1984).

Figure Captions

Figure 1- Inverse magnetic susceptibility $\chi^{-1}$ vs $T$ of UNi$_{0.5}$Sb$_2$, for $\mu_0 H = 1$ T. A fit of these data to a Curie-Weiss law for 250 K $< T <$ 350 K yielded the effective moments of $\mu_{eff} \approx$ 3.15 ($H$//c) and 3.25 $\mu_B$ ($H \perp$c). The inset shows the behavior of $\chi$ vs $T$ below 100 K, detailing the difference between the cooling and warming cycles of the features centered near 40 and 85 K.

Figure 2- Specific heat $C_p$ vs $T$ for UNi$_{0.5}$Sb$_2$. The peak at $T_N \approx$ 161 K is consistent with the onset of antiferromagnetic order. The inset shows a plot of $C_p/T$ vs $T^2$ at low temperatures; an extrapolation of these data to $T = 0$ yields a Sommerfeld coefficient $\gamma =$ 9.5 mJ/mole·K

Figure 3- Normalized electrical resistivity $\rho/\rho_{300K}$ vs $T$ in UNi$_{0.5}$Sb$_2$ for $P_{300K}$ ($P_{T<90K}$) = 0 (0), 5.2 (1.9), 11.0 (9.0), 15.4 (14.8), and 17.8 (20.6) kbar. The curves at the different pressures are offset for clarity. The arrows near the low temperature features for $P = 0$ differentiate between cooling and warming cycles. Pressure broadens and lowers the onset temperature of the features near 40 and 85 K. At $P_{300K} = 15.4$ kbar ($P_{T<90K} = 14.8$ kbar) the 2 features start to merge and to be suppressed, and they cannot be distinguished anymore for $P_{300K} = 17.8$ kbar ($P_{T<90K} = 20.6$ kbar). The inset shows that $T_N$ increases with pressure at the approximate rate of 0.76 K/kbar. The pressure values in the main pane are values at 300 K. The pressure values in the inset are estimated (see text).

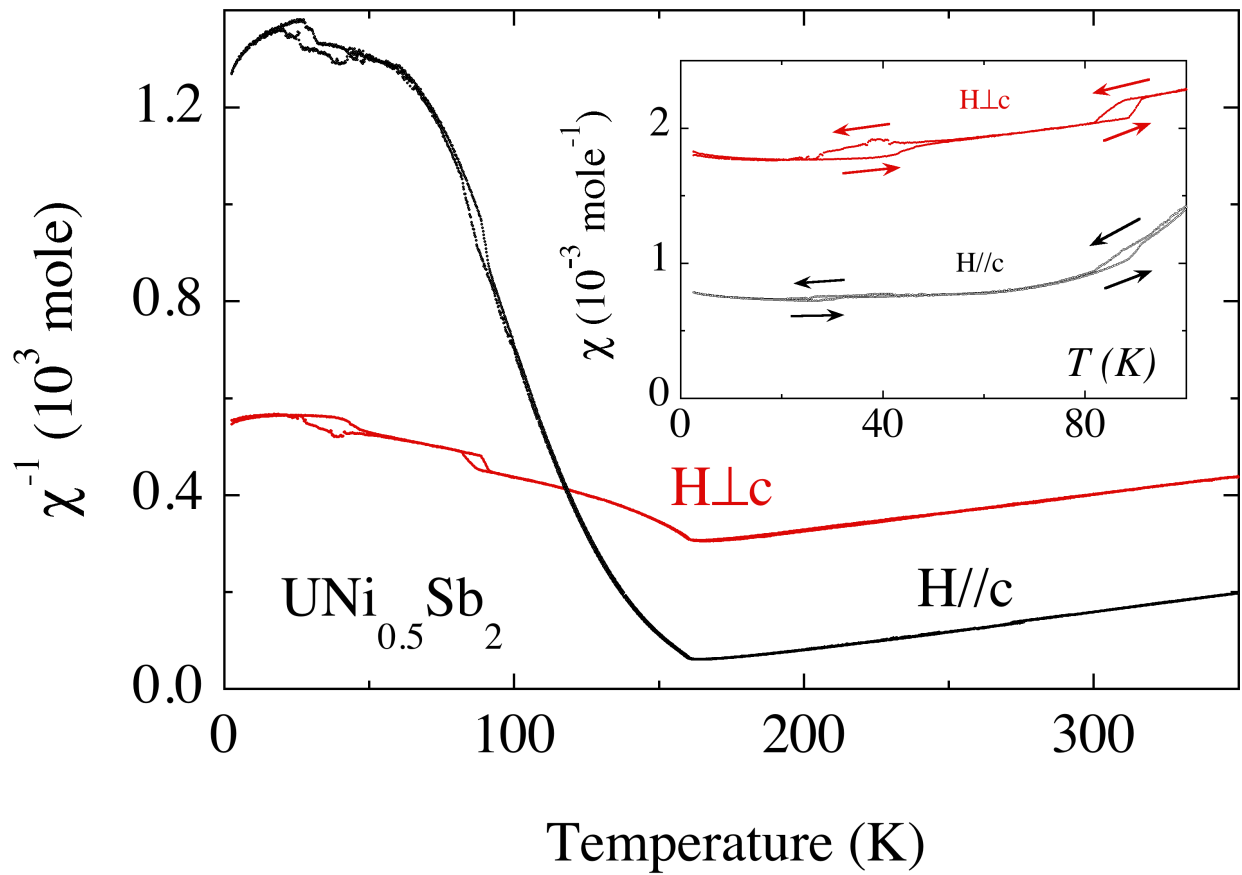

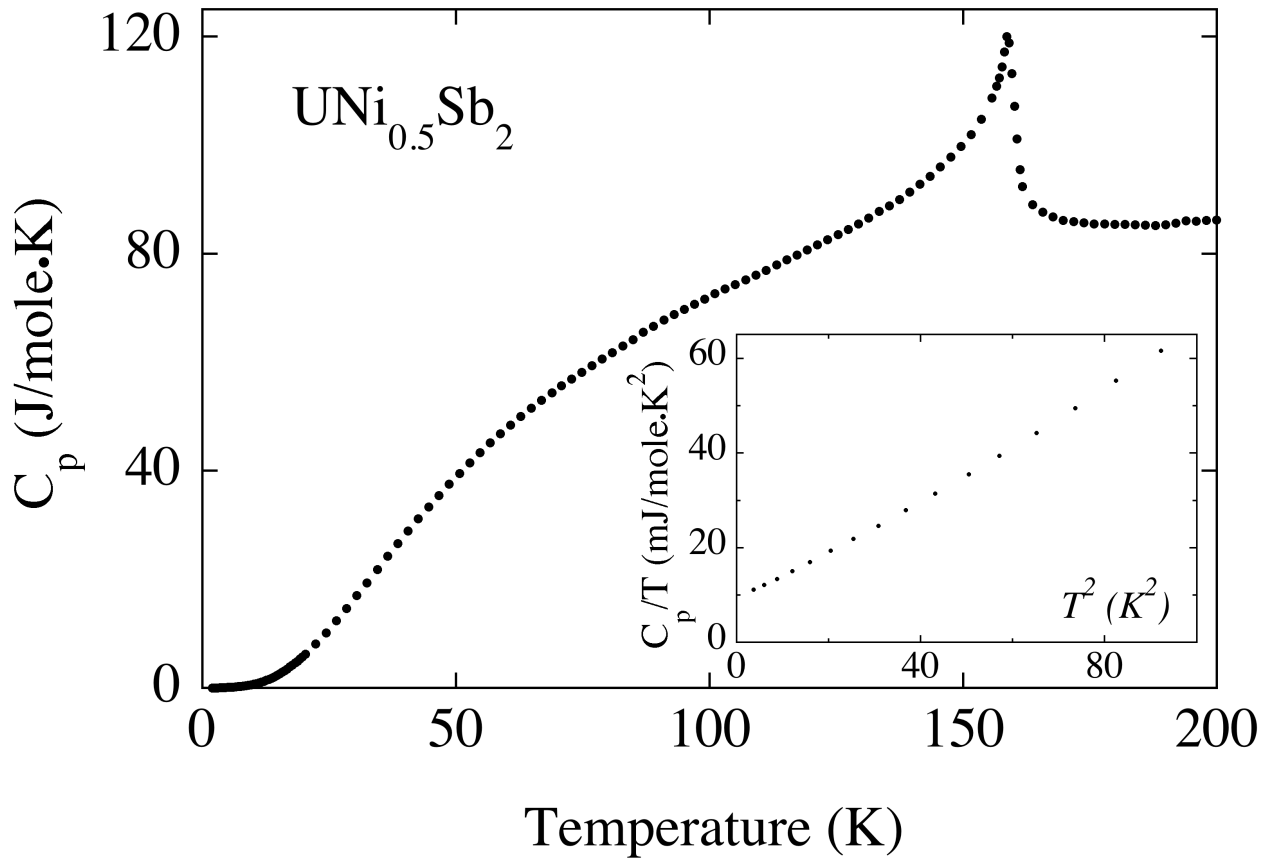

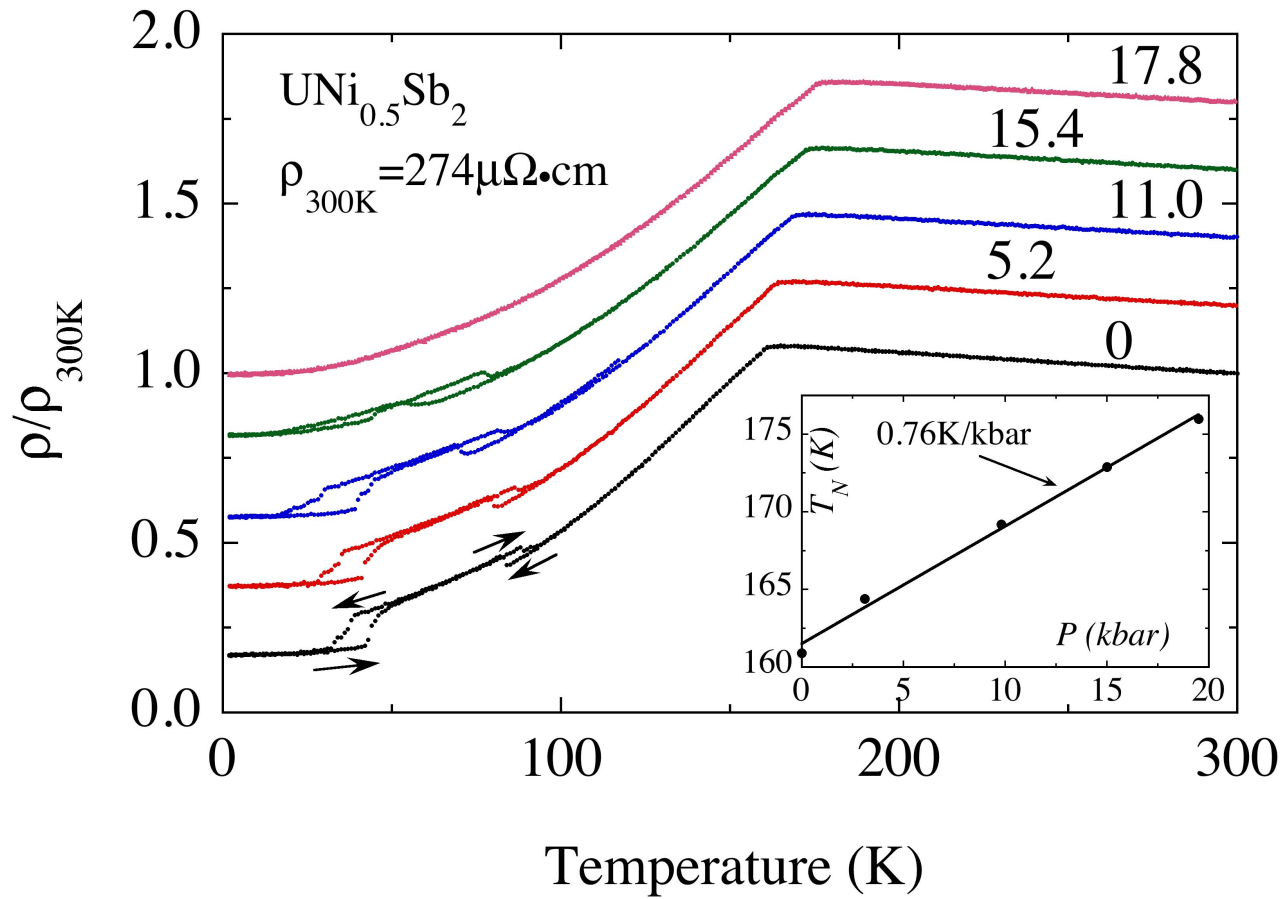